\documentclass{emulateapj}

\shorttitle{Kepler's Supernova Remnant}

\shortauthors{Patnaude et al.}

\begin{document}

\title{The Origin of Kepler's Supernova Remnant}

\author{Daniel J.~Patnaude\altaffilmark{1},
Carles Badenes\altaffilmark{2,3,4}, Sangwook Park\altaffilmark{5}, and 
J.~Martin Laming\altaffilmark{6}}

\altaffiltext{1}{Smithsonian Astrophysical Observatory, Cambridge, 
MA 02138, USA}

\altaffiltext{2}{University of Pittsburgh, 409 Allen Hall, 
Pittsburgh, PA 15260}

\altaffiltext{3}{School of Physics and Astronomy, Tel-Aviv University,
Tel-Aviv 69978, Israel}

\altaffiltext{4}{Benoziyo Center for Astrophysics, Weizmann Institute
of Science, Rehovot 76100, Israel}

\altaffiltext{5}{Department of Physics, University of Texas at Arlington,
Arlington, TX 76019}

\altaffiltext{6}{Space Science Division, Naval Research Laboratory, Code 
7674L, Washington, DC 20375, USA}

\begin{abstract}

It is now well established that Kepler's supernova remnant is the result of 
a Type Ia explosion. With an age of 407 years, and an angular diameter
of $\sim$ 4$\arcmin$, Kepler is estimated to be between 3.0 and 7.0 kpc 
distant. 
Unlike other Galactic Type Ia supernova remnants such as Tycho and SN~1006, and
SNR 0509-67.5 in the Large Magellanic Cloud, 
Kepler shows evidence for a strong circumstellar interaction. A bowshock 
structure in the north is thought to originate from the motion of a 
mass--losing system through the interstellar medium prior to the supernova. 
We present results of hydrodynamical and spectral modeling aimed at 
constraining the circumstellar environment of the system and the 
amount of $^{56}$Ni produced in the explosion. Using models that contain
either 0.3M$_{\sun}$ (subenergetic) or 1.0M$_{\sun}$ (energetic) of $^{56}$Ni,
we simulate the interaction between supernova Ia ejecta 
and various circumstellar density models. Based on dynamical
considerations alone, we find that the subenergetic models favor a distance 
to the SNR of $<$ 6.4 kpc, while the model that produces
1M$_{\sun}$ of $^{56}$Ni requires a distance to the SNR of $>$ 7 kpc.  
The X-ray spectrum is consistent
with an explosion that produced $\sim$ 1M$_{\sun}$ of $^{56}$Ni, ruling out
the subenergetic models, and suggesting that Kepler's SNR was a SN~1991T-like
event. Additionally, the
X-ray spectrum rules out a pure $r^{-2}$ wind profile expected from isotropic
mass loss up to the time of the supernova. Introducing a small cavity
around the progenitor system results in modeled X-ray spectra that are 
consistent with the observed spectrum. If a wind shaped circumstellar
environment is necessary to explain
the dynamics and X-ray emission from the shocked ejecta in Kepler's SNR, then
we require that the distance to the remnant be greater than 7 kpc.

\end{abstract}


\section{Introduction}

Type Ia supernovae (SNe) are believed to be the thermonuclear explosion of 
a C+O white dwarf (WD) that is destabilized when its mass approaches the 
Chandrasekhar limit by accretion of material from a binary companion. After 
the central regions ignite, a burning front propagates outwards, consuming 
the entire star and leading to a characteristic ejecta structure with 
$\sim$ 0.7 M$_{\sun}$ of Fe-peak nuclei (primarily in the form of $^{56}$Ni 
that powers the SN light curve) in the inside, and an equal amount of 
intermediate mass elements on the outside. While the amount of $^{56}$Ni can 
vary between $\sim$ 0.3 M$_{\sun}$ and $\sim$ 1 M$_{\sun}$, the 
stratification of the ejecta remains constant across SNe Ia subtypes 
\citep{mazzali07}. The simple structure of the ejecta results in uniform 
light curves and spectra and make SN~Ia useful distance indicators
\citep[e.g.,][]{riess98,perlmutter99}. 
Depending upon the nature of the WD companion, SN~Ia progenitors are divided 
into either single degenerate \citep{hachisu96} or double degenerate cases 
\citep{iben84,webbink84}. Finding clear evidence supporting one of these two 
scenarios has become a cornerstone in stellar astrophysics, as unknown 
evolutionary effects associated with the progenitors may introduce systematic 
trends that could limit the precision of cosmological measurements based on 
SN~Ia \citep{howell09}.

Most Type Ia supernova remnants (SNRs) reflect the simplicity of the 
explosion. Galactic SNRs such as Tycho and SN~1006 appear remarkably 
symmetric \citep{warren05,chenai08}, unlike their core--collapse 
counterparts, such as Cas A \citep{delaney10,lopez11,hwang12} or younger 
extragalactic SNe such as SN~1980K or SN~1993J \citep{milisavljevic12}. 
The morphology and X-ray spectra of SNRs with 
known ages and a Type Ia classification also indicate that the progenitors 
do not modify their surroundings in a strong way -- in particular, while absorption
features seen in the spectra of Type Ia SNe suggest small cavities of radius
$\gtrsim$ 10$^{17}$ cm \citep{patat07,blondin09,borkowski09,simon09,sternberg11}, there is 
little evidence for large ($r$ $\sim$ 3 -- 30 pc) 
wind-blown cavities around the explosion site 
\citep{badenes07}. However, \citet{williams11} recently showed that the
candidate Type Ia RCW~86 is currently interacting with a cavity wall and that
the cavity has a radius of $\sim$ 12 pc, in the middle of the range
of sizes expected from the accretion wind models \citep{badenes07}. 
The lack of a sizeable population of Ia SNR inside of cavities 
is particularly relevant as the current paradigm for 
SN~Ia progenitors in the single degenerate channel predicts fast, optically 
thick outflows from the WD surface which would leave behind such cavities 
\citep[e.g.][]{hachisu96}. Additionally, the dynamical and spectral properties 
of young Ia SNRs like Tycho and 0509--67.5 are more consistent with an 
interaction with a constant density interstellar medium (ISM) 
\citep{badenes06,badenes08}.

\begin{figure}
\plotone{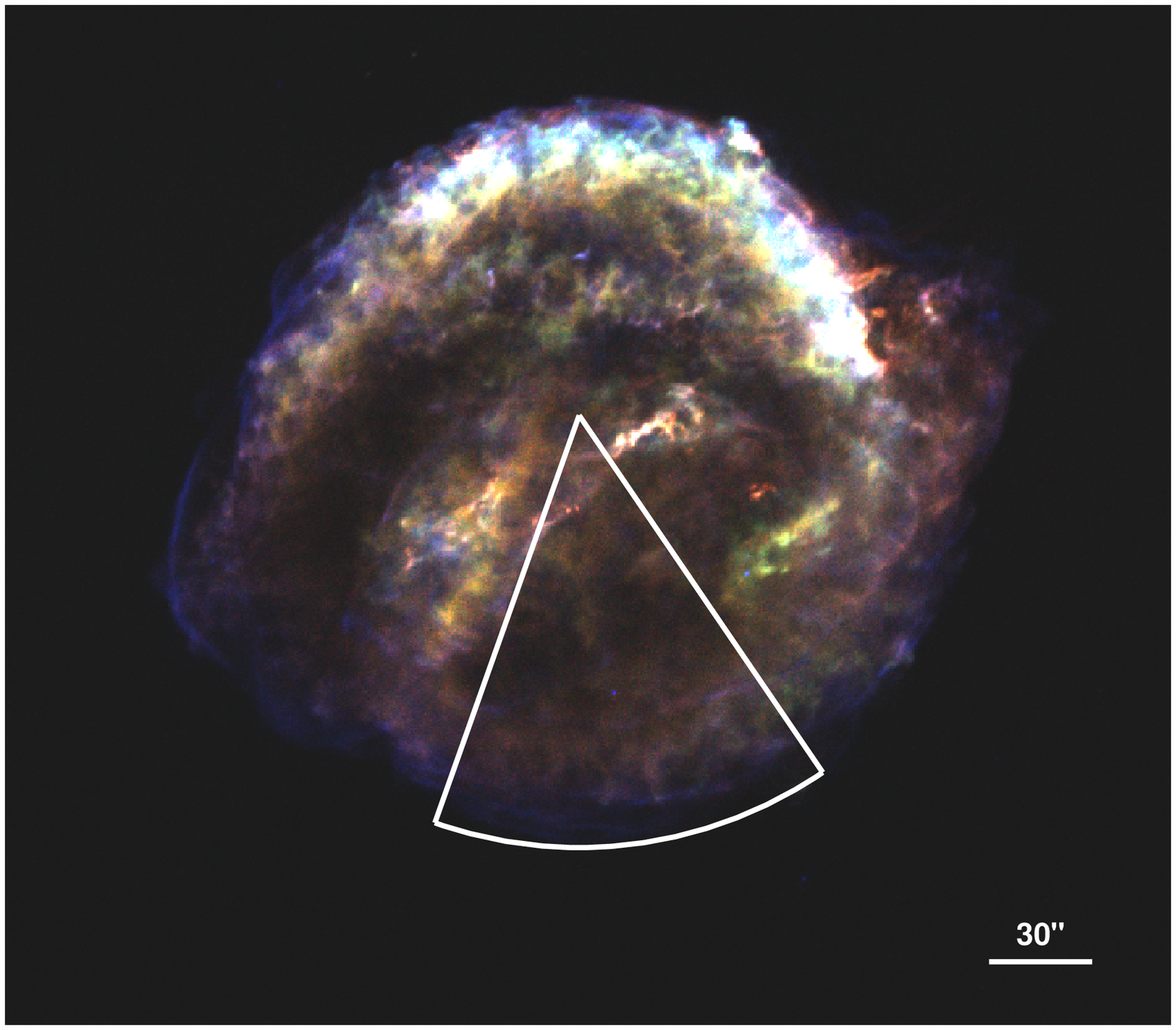}
\caption{Kepler's SNR viewed in X-rays with {\it Chandra}
ACIS-S3. The RGB image shows $0.4$--$0.75$ keV emission in red, $0.75$--$1.2$ keV
emission in green, and $1.2$--$7.0$ keV emission in blue. In the image, north is up and east is to the left.}
\label{fig:kepler}
\end{figure}

A notable exception to this is Kepler's SNR (G4.5+6.8; Kepler), shown in X-rays in
Figure~\ref{fig:kepler}. Kepler's 
SNR has recently been firmly established as the remnant of a Type Ia SN, based
on the O/Fe ratio observed in the X-ray spectrum \citep{reynolds07}. However, 
unlike many other Type Ia SNRs, the morphology of this $\approx$ 400 yr old
remnant shows a strong bilateral
symmetry, and optical observations reveal clear signs of an interaction
between the blastwave and a dense, nitrogen--enhanced circumstellar
shell with a mass of $\sim$ 1M$_{\sun}$ \citep{vink08}. Near infrared
observations also show evidence for circumstellar knots akin to the
quasi-stationary floculi of Cassiopeia A \citep{gerardy01}.
This suggests that either the 
progenitor of Kepler's SN or its binary companion might have been relatively 
massive, creating a bowshock shaped circumstellar medium (CSM) 
structure as the system lost mass 
and moved against the surrounding ISM. \citet{borkowski92} found values
of $\dot{M}$ $\sim$ 5$\times$ 10$^{-6}$ M$_{\sun}$ yr$^{-1}$ and 
v$_{wind}$ = 15 km s$^{-1}$ are required in order to explain the
morphology of Kepler's bowshock. While it is likely that not all 
progenitor scenarios for SNe Ia are the same, the dense CSM wind required
to create the observed morphology appears to be in conflict with
observations of the recent nearby SN~Ia 2011fe in M101 \citep{nugent11,li11}. 
In the case of 2011fe, the radio 
and X-ray observations place constraints on the circumstellar environment
with $n_{\mathrm{CSM}}$ $\lesssim$ 6 cm$^{-3}$, or in the case of 
a stellar wind, $\dot{M}$ $\lesssim$ 6$\times$10$^{-10}$ M$_{\sun}$ yr$^{-1}$
\citep{horesh12,chomiuk12}. 
It is worth noting, however, that the recent observations of SN~2011fe 
probe spatial scales of $\sim$ 10$^{16}$ cm, corresponding to the
last $\sim$ 100--1000 yr of the progenitor's evolution, while the blastwave
of Kepler's SNR is probing the mass-loss history 10$^5$ -- 10$^{6}$ yr ago.

Located well above the Galactic plane, the SNR has an angular size of
2$\arcmin$ in radius, but the distance is poorly constrained.
\citet{reynoso99} used \ion{H}{1} absorption to estimate a distance
of 4.8 $<$ D$_{\mathrm{SNR}}$ $<$ 6.4 kpc, while \citet{sankrit05} 
estimate that the
SNR is somewhat closer, with a lower limit of 3.0 kpc on the distance, 
based on the width of the H$\alpha$ line and proper motion measurements.
The non-detection of TeV $\gamma$-rays by HESS (F$_{\gamma}$ $<$ 8.6 $\times$
10$^{-13}$ erg cm$^{-2}$ s$^{-1}$) places a lower limit on the 
distance to Kepler of $\sim$ 6 kpc \citep{aharonian08} 
assuming a model for the gamma-ray
emission \citep{berezhko06}, for a normal Ia SN. To match the
angular size, the distance would have to be even greater than 6 kpc 
($\gtrsim$ 7 kpc) if the SN were more energetic than a normal Ia, or 
else the blastwave would have to expand into a CSM with considerable
density.
\citet{vink08} measured the velocity of the SNR forward shock
and estimated $v_s$ = 4200$d_{4\mathrm{kpc}}$ km s$^{-1}$ and concluded
that the distance must be $\gtrsim$ 6 kpc or otherwise the SN was 
subenergetic. 
The distance estimates are incompatible with one another: either 
3.0 $<$ D$_{\mathrm{SNR}}$ $<$ 6.4 kpc, or D$_{\mathrm{SNR}}$ $>$ 7 kpc, 
based on the non-detection of TeV gamma-rays, an important new constraint.
Given the uncertainty on the distance, the blastwave has a radius
between 1.75 $\geq$ R$_{FS}$ $\geq$ 3.7 pc, or R$_{FS}$ $>$ 4.1 pc.

Here we present hydrodynamical simulations of Type Ia SN ejecta interacting 
with an external medium shaped by a circumstellar wind. Combining 
models for the pre-supernova mass-loss 
with models for the ejecta density distribution and composition, we 
simultaneously fit both the morphology of the SNR and the observed
 X-ray spectrum within the observable constraints-- the age and angular 
size of the SNR along with 
the bulk spectral properties. Since the circumstellar 
environment is complicated by the bowshock structure in the north, we focus 
our attention on the southern portion of the SNR (see Fig.~\ref{fig:kepler}), 
in an attempt to minimize 
the number of free parameters in our hydrodynamical simulations. Based
on our joint hydrodynamical and spectral fits, we conclude that Kepler was a 
luminous Type Ia SN. If the SNR is expanding into a stellar wind we require a
distance to Kepler of $>$ 7 kpc. If, on the other hand, the SNR is expanding
into a uniform medium, the distance is somewhat closer, $\sim$ 5 kpc.

\section{X-ray Emission from Kepler's SNR}

\begin{figure}
\plottwo{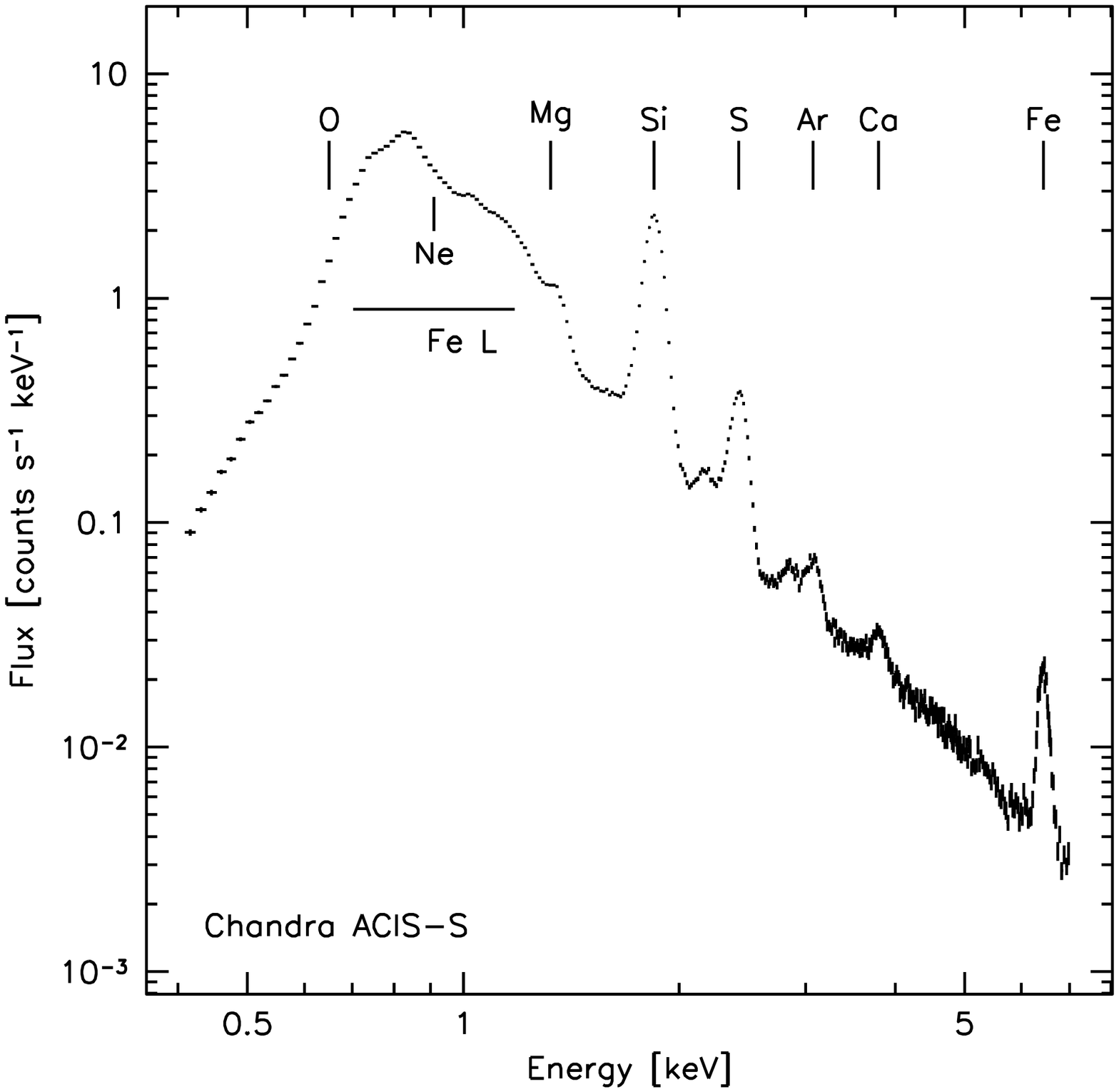}{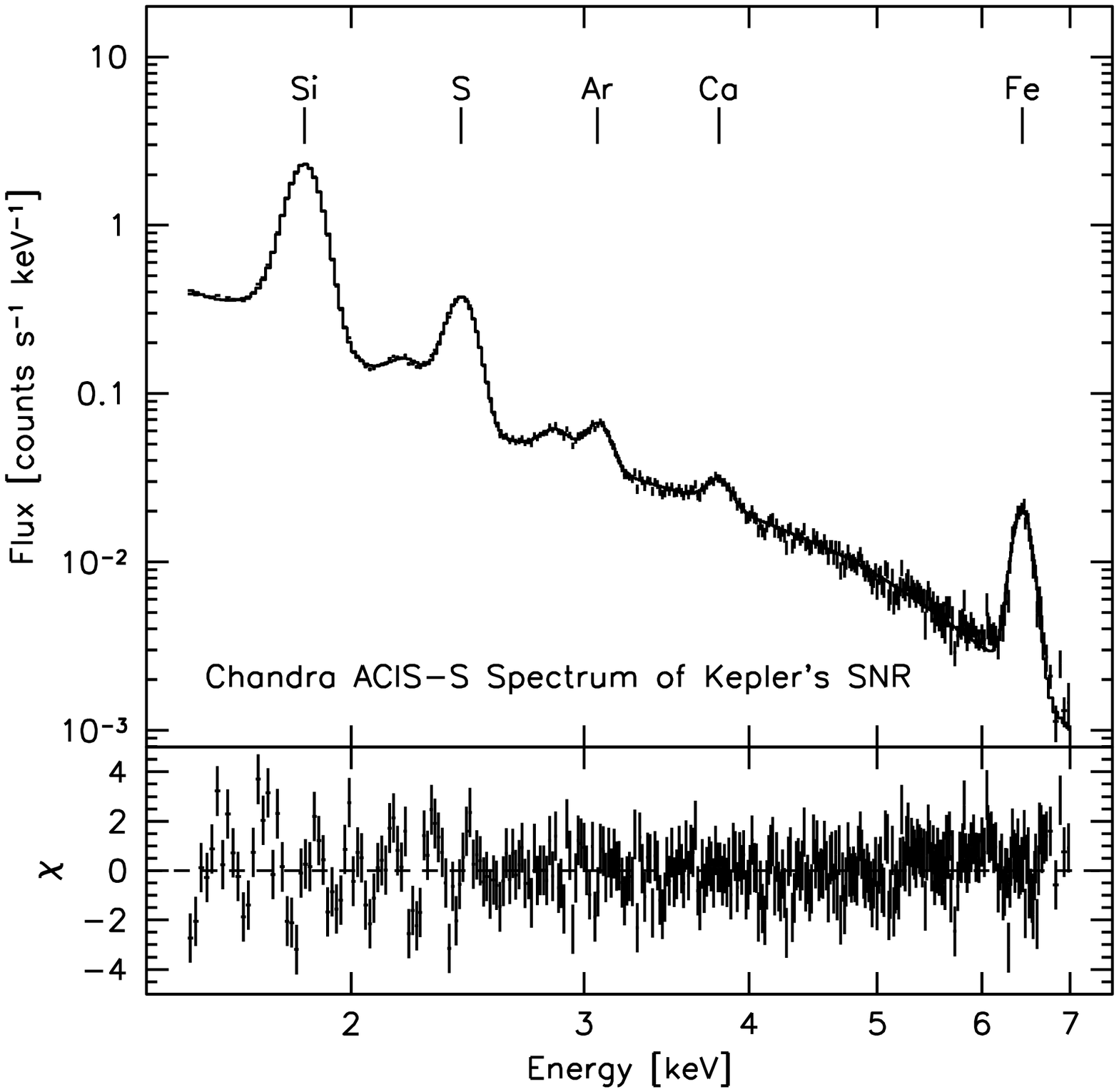}
\caption{{\it Left}: {\it Chandra} ACIS-S spectrum of the pie-sliced region
of Kepler's SNR shown in Figure~\ref{fig:kepler}. For reference, we mark the
positions of several K$\alpha$ lines that are used in the analysis. {\it Right}: Spectral
fit to the 1.7--7.0 keV spectrum of the pie-sliced region shown in 
Figure~\ref{fig:kepler}.The fit parameters are listed in 
Table~\ref{tab:fits}, and the positions of the K$\alpha$ lines are also marked. In 
both figures, the data are unbinned (1024 PHA channels).}
\label{fig:spec_fit}
\end{figure}

A 750ks {\it Chandra} image of Kepler's SNR (PI: S.~Reynolds) 
is shown in Figure~\ref{fig:kepler}, with the 
X-ray emission from the pie-sliced region shown in 
Figure~\ref{fig:spec_fit} (left). As shown in Fig~\ref{fig:spec_fit} (left),
the X-ray spectrum is dominated by emission from Fe K and Fe L, as well as 
emission from intermediate mass elements such as silicon, sulfur, and argon. 
The amount of iron emission in the spectrum is a key constraint on the explosion, as 
it will allow us to discriminate between Type Ia explosion models 
\citep{badenes05}. 
Besides emission from the shocked ejecta, there is also emission from 
circumstellar material-- mainly oxygen and neon, which overlaps with the 
L-shell iron emission \citep{reynolds07}, 
as well as nonthermal continuum which contributes to 
the emission above a few keV \citep{reynolds07,chenai08}. 
In the subsequent discussion on the fits to 
the spectrum, we do not account for either the contribution from the 
swept up circumstellar material or nonthermal emission in the spectral models-- 
the contribution by
shocked circumstellar material to the thermal X-ray emission is small and
accounts for only a few percent of the total emission \citep{reynolds07}, 
while the nonthermal emission serves to raise the continuum above $\sim$ 4 keV.
The processes that lead to the nonthermal emission could alter the thermal
spectrum \citep{patnaude09,patnaude10}, but those affects have been shown
to be small, particularly in young SNRs \citep{patnaude10}. Interesting
characteristics of the spectrum include: the presence or absence of 
emission from iron, which will constrain the explosion energetics; the
centroids of K--shell emission lines, which constrains the ionization 
state of the shocked gas; and flux ratios of individual lines, such as
K$\alpha$ to H-like Ly$\alpha$, which constrain the density of the circumstellar
material. 

The spectra from the combined 750ks observation  
shown in Figure~\ref{fig:kepler} was extracted from the
Level 2 {\it Chandra} event list using CIAO 4.1, and weighted response
matrices were computed using CalDB version 4.1.3. 
The broadband spectrum (Fig.~\ref{fig:spec_fit}, left) 
shows significant emission from both Fe L and Fe K \citep{reynolds07}. Given
the complexity of the Fe L emission at $\sim$ 1 keV, we focus only
on those emission lines above 1.5 keV, as they are well separated from neighboring
lines. 
To measure the emission line centroids, we modeled the spectrum 
as a powerlaw continuum with a series of Gaussians for each emission line. The
spectral fit is shown in Figure~\ref{fig:spec_fit} (right). 
We chose to model the continuum as a powerlaw since \citet{chenai08} showed
that much of the continuum emission in Kepler is nonthermal in 
nature \citep[see Figs.~6 and 8 of][]{chenai08}. For
the spectral fits, we used XSPEC Version 11.3.2ag. We used a region
nearby the supernova remnant for a local background subtraction. Since
we are only fitting the data above $\sim$ 1.5 keV, we do not include
absorption in our spectral fits. \citet{reynolds07} found an absorbing
column of N$_H$ = 5.2 $\times$ 10$^{21}$ cm$^{-2}$ by fitting featureless
regions along the remnant's rim to an absorbed {\tt srcut} model. 
Fitting only the data above
1.5 keV, we were unable to constrain the column density. However in the
subsequent sections where we compare our hydrodynamical and spectral 
models to the data, we
assume a column of 5.2 $\times$ 10$^{21}$ cm$^{-2}$ so that we may
compare the computed X-ray emission below 1.5 keV directly to the data.

The results of our 
centroid fits are listed in Table~\ref{tab:fits}. The key results in 
Table~\ref{tab:fits} are the line centroids, and in the subsequent 
discussion, we will focus on the line centroids of 
of Si K$\alpha$, S K$\alpha$, and Fe K$\alpha$. We also point out in
Figure~\ref{fig:spec_fit} (right) that there is little emission at 2 keV from 
\ion{Si}{14} Ly$\alpha$. This is confirmed in our spectral analysis where we 
are only able to set an upper limit on the \ion{Si}{14} Ly$\alpha$ emission
from shocked ejecta in the south. 
The lack of emission at this energy will be used to constrain
the CSM structure and density.

\section{Hydrodynamical Modeling}

In this section, we present results from a modeling effort where we
have coupled the hydrodynamics to a nonequilibrium ionization (NEI) 
calculation to produce joint 
spectral and dynamical fits to Kepler's SNR. We describe, in general terms
in \S~3.1 the formation of the bowshock structure in the north of Kepler and how it can be 
used to constrain the mass-loss parameters of the progenitor system. We also 
describe how the choice of Ia explosion model (e.g. the explosion energy and amount 
of $^{56}$Ni that is synthesized) imposes a constraint on the uncertain
distance to the SNR. In the subsequent 
sections, we compare the bulk spectral properties of Kepler's SNR to those
predicted by our spectral models.

\subsection{Morphology of Kepler's SNR in the Context of a CSM Interaction}

The presence of a bowshock and shell of circumstellar material 
in the north of Kepler's SNR suggests that
there was an outflow from the progenitor or its companion prior to the
explosion. This is because the bowshock can only form at a converging flow, 
in this case a 
wind interacting with the interstellar medium, where the ISM has a net inflow
velocity relative to the progenitor, due to the progenitor's systemic
velocity through the ISM. \citet{chiotellis12} recently modeled the dynamics
of the SNR blastwave through the bowshock and concluded that the
dynamics of the SNR blastwave are consistent with the presence of
a bowshock. We can use
estimates for the location of the bowshock stagnation point as a 
constraint on the wind parameters. Generally, 
the location of the stagnation point, $r_0$, is written as 
\citep[e.g.][]{huang82,borkowski92}:

\begin{equation}
r_0 = 1.78\times10^{3}\left(\frac{\dot{M} v_{\mathrm{wind}}}{n_{\mathrm{amb}} u_{\star}^2}\right)^{1/2} \mathrm{pc} \ ,
\label{eqn:stag}
\end{equation}

\noindent
where $\dot{M}$ is the mass-loss rate, $v_{\mathrm{wind}}$ is the wind speed, 
$n_{\mathrm{amb}}$ is the ambient ISM density, and $u_{\star}$ is the 
systemic velocity of the star. In order to explain the large distance above 
the Galactic plane, the progenitor likely had a large systemic velocity, 
$u_{\star}$ $\sim$ 250 km s$^{-1}$ \citep{bandiera87}, a notion confirmed 
by the high proper motion and radial velocities of nitrogen-rich knots 
in the remnant, as well as the narrow component of the H$\alpha$ line in the 
nonradiative shock. \citet{borkowski92} estimated that the undisturbed
ISM material had a density of 10$^{-3}$ -- 10$^{-4}$ cm$^{-3}$, 
corresponding to the hot component of the ISM \citep{mckee77}. 
Equation~\ref{eqn:stag}
can be inverted to estimate the mass-loss rate of the wind:

\begin{equation}
\dot{M} = 3.15\times 10^{-7} \frac{n_{\mathrm{amb}} u_{\star}^2}{v_{\mathrm{wind}}} \left(\frac{r_0}{\mathrm{pc}}\right)^2 \ .
\label{eqn:mdot}
\end{equation}

Given the uncertainty of greater than a factor of two on the distance, the 
stagnation point is located at a radius of 2--4 pc, similar to the 
2--3 pc assumed by \citet{borkowski92}. The amount of mass in 
the shell is $\approx$ 1 M$_{\sun}$. \citet{chiotellis12} have suggested that 
the nitrogen--rich knots in the CSM, combined with the solar mass of material 
in the shell, argue in favor of the CSM being sculpted by the outflow from 
an asymptotic giant branch (AGB) star. The mass-loss rates from AGB stars can 
vary over several orders of magnitude, but the velocity of the wind is 
generally 10--20 km s$^{-1}$ \citep{schoier01,olofsson02,ramstedt09}. 
Using this range in velocity combined with the 
range in the radius of the stagnation point yields mass-loss rates of 
10$^{-6}$ -- 10$^{-5}$ M$_{\odot}$ yr$^{-1}$. 

To model a CSM wind, the mass-loss rate and wind velocity are all that are 
required, since $\rho_{\mathrm{amb}}$ = $A r^{-2}$ where the wind normalization 
$A$ $=$ $\dot{M}$/(4$\pi$ $v_{\mathrm{wind}}$). We couple this model for the 
circumstellar environment to models for Type Ia SN ejecta from \citet{badenes03}. 
We choose to focus on models DDTa (E$_{\mathrm{SN}}$ = 1.4$\times$10$^{51}$ 
erg) and DDTg (E$_{\mathrm{SN}}$ = 0.9$\times$10$^{51}$ erg). The choice
of models is designed to encompass the diversity of Ia explosions
\citep{badenes08}. In a constant 
density environment, the choice of explosion model does not significantly 
influence the radius and velocity of the forward shock as long as the kinetic 
energy is $\sim$ 10$^{51}$ erg. In a wind environment, the difference in 
forward shock radius between an energetic and subenergetic model, where
the ejecta is approximated as a powerlaw in velocity, is a little 
larger, since R$_{\mathrm{FS}}$ $\propto$ E$_{\mathrm{SN}}^{2/(n-2)}$ 
rather than E$_{\mathrm{SN}}^{2/n}$ \citep{chevalier82}. Of more 
interest is the resultant X-ray emission, which is influenced by the 
density and chemical composition of the ejecta. For instance, in the DDTa 
model, $\sim$ 1M$_{\sun}$ of $^{56}$Ni is produced, compared to 0.3M$_{\sun}$ in 
the DDTg model. Since the $^{56}$Ni decay is responsible for the iron
observed in the ejecta, the amount of $^{56}$Ni produced in the SN 
will profoundly affect the emitted X-ray spectrum.

\begin{figure}
\plotone{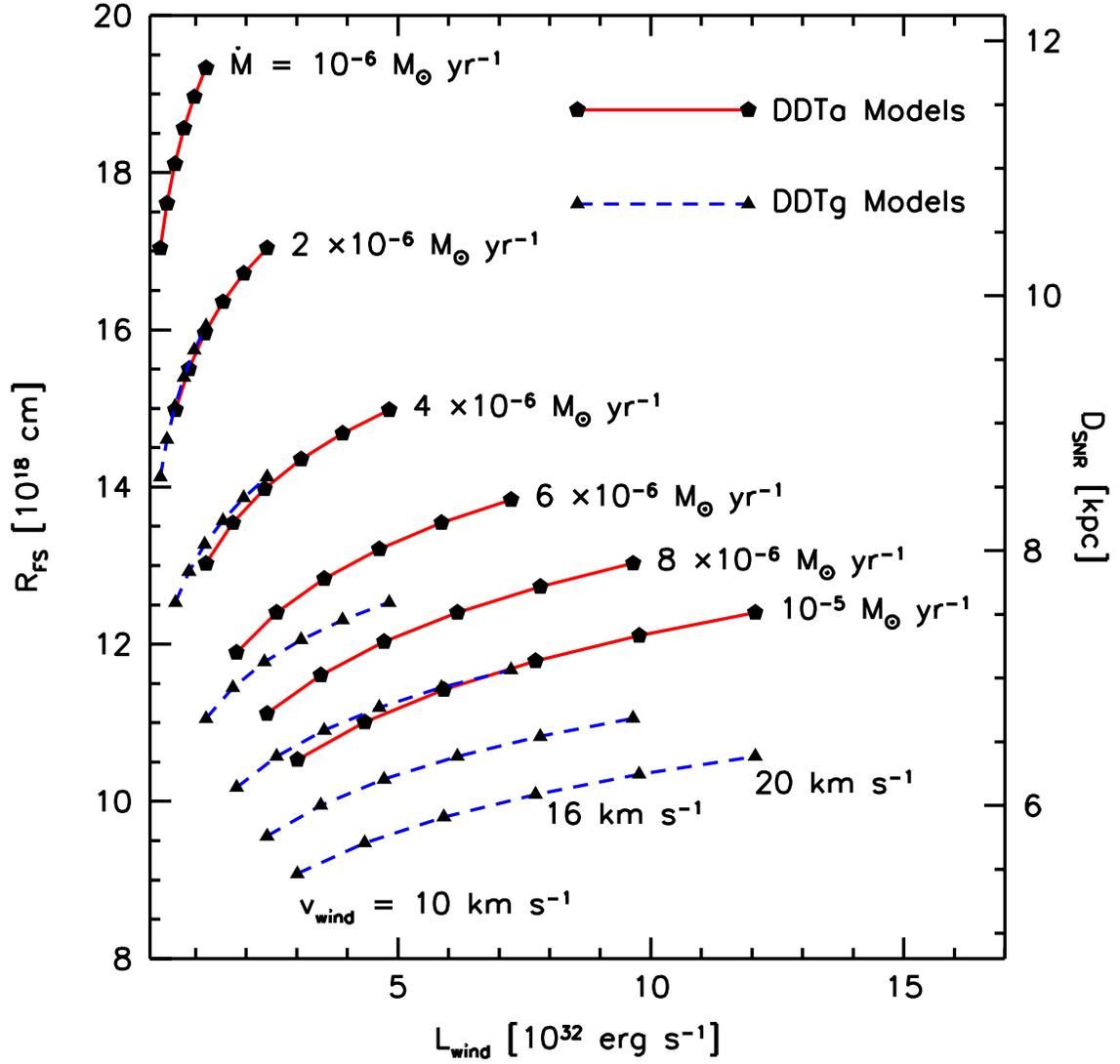}
\caption{Forward shock radius as a function of wind luminosity for a range of mass-loss 
rates and wind speeds, for Type Ia SN ejecta models DDTa and DDTg. On the right
hand axis is the distance to Kepler that is implied by the forward shock radius, 
for an angular diameter of 4$\arcmin$, but the models may be scaled to any distance.}
\label{fig:bounds}
\end{figure}

Using the VH-1 hydrodynamics code, a numerical hydrodynamics code developed
at the University of Virginia by J.~Hawley, J.~Blondin, and 
collaborators \citep[e.g.,][]{blondin93}, we modeled the expansion of the 
SN ejecta 
into the CSM wind to an age of 400 yr. We varied the wind speed between 
10--20 km s$^{-1}$ and the mass-loss rate as described above. The parameter 
space results in a grid of models that relate the radius of the forward shock 
with the wind mechanical luminosity, $L_{\mathrm{wind}}$ = 
$\frac{1}{2}\dot{M}v_{\mathrm{wind}}^2$, for each explosion model. In 
Figure~\ref{fig:bounds}, we plot the shock radius as a function of wind 
luminosity for each combination of wind speed and mass-loss rate.
Roughly speaking $R_{\mathrm{FS}}$ $\propto$ $L_{\mathrm{wind}}^{1/5}$ $\propto$
v$_{\mathrm{wind}}^{2/5}$, for a given mass-loss rate. 
This is because the CSM density normalization 
is set by the wind speed -- higher wind velocities transport the mass further 
away from the progenitor, while lower velocity winds result in more mass 
deposited closer 
to the progenitor. In the higher wind velocity models, the blastwave 
propagates further after 400 years since it has to move through less 
material. 

As discussed earlier, the distance to Kepler is not well known, ranging
from 3 -- 6.4 kpc, or $>$ 7 kpc. In Figure~\ref{fig:bounds}, 
we also note the distance
to Kepler on the right hand axis, using the measured angular radius of 2$\arcmin$. 
The upper limit
of 6.4 kpc \citep{sankrit05} on the distance rules out {\it all} DDTa models on dynamical
considerations alone. However, if the distance is $>$ 7 kpc, as expected
from the TeV gamma-ray non-detection, then most DDTa models are allowed, while
a sizeable fraction of DDTg models ($\dot{M}$ $\gtrsim$ 6$\times$ 10$^{-6}$ M$_{\sun}$ yr$^{-1}$) 
are no longer favored.

\subsection{Spectral Fitting: Wind Models}

To compare our grid of dynamical models to the observations of Kepler, we have
generated synthetic X-ray spectra for the emission from shocked ejecta in our 
hydrodynamic simulations following the methods presented in 
\citep{badenes03,badenes05}, using updated atomic data \citep{badenes06},
and including radiative and ionization losses as described in 
\citet{badenes07}. In a constant density ambient medium, the synthetic spectra
are controlled by three variables: the density of the ambient material,
the SNR age, and the amount of collisionless heating between electrons
and ions ($\beta$) at the reverse shock. For our models where 
$\rho_{\mathrm{amb}}$ $\propto$ $r^{-2}$, the
X-ray emission is instead influenced by the wind normalization (which sets
the density). Higher values of $\dot{M}$/(4$\pi$ $v_{\mathrm{wind}}$) will
result in more circumstellar material closer to the explosion. 
We expect that this will
lead to more collisional ionization in the shocked ejecta which will be 
directly reflected in the X-ray spectrum.

The ability of our synthetic spectra to reproduce the observations is limited
by the quality of the atomic data used in the code. The data for 
K$\alpha$ blends are reasonably complete however, and we thus focus on 
those lines. As shown in Table~\ref{tab:fits}, we have measured the line
centroids and fluxes of several K$\alpha$ lines. These can be compared
directly against the synthetic spectra from our models. We focus only
on the brightest lines of silicon, sulfur, and iron, and note the following
additional constraints on our models-- (1): the presence of Fe K emission at 
$\sim$ 6.4 keV; (2): the absence of significant \ion{Si}{14} Ly$\alpha$ emission
at $\sim$ 2.0 keV; (3): the amount of Fe L emission  below 1 keV, relative
to oxygen suggests that the emission there is in fact ejecta and not
shocked circumstellar material \citep{reynolds07}.

\begin{figure}
\plottwo{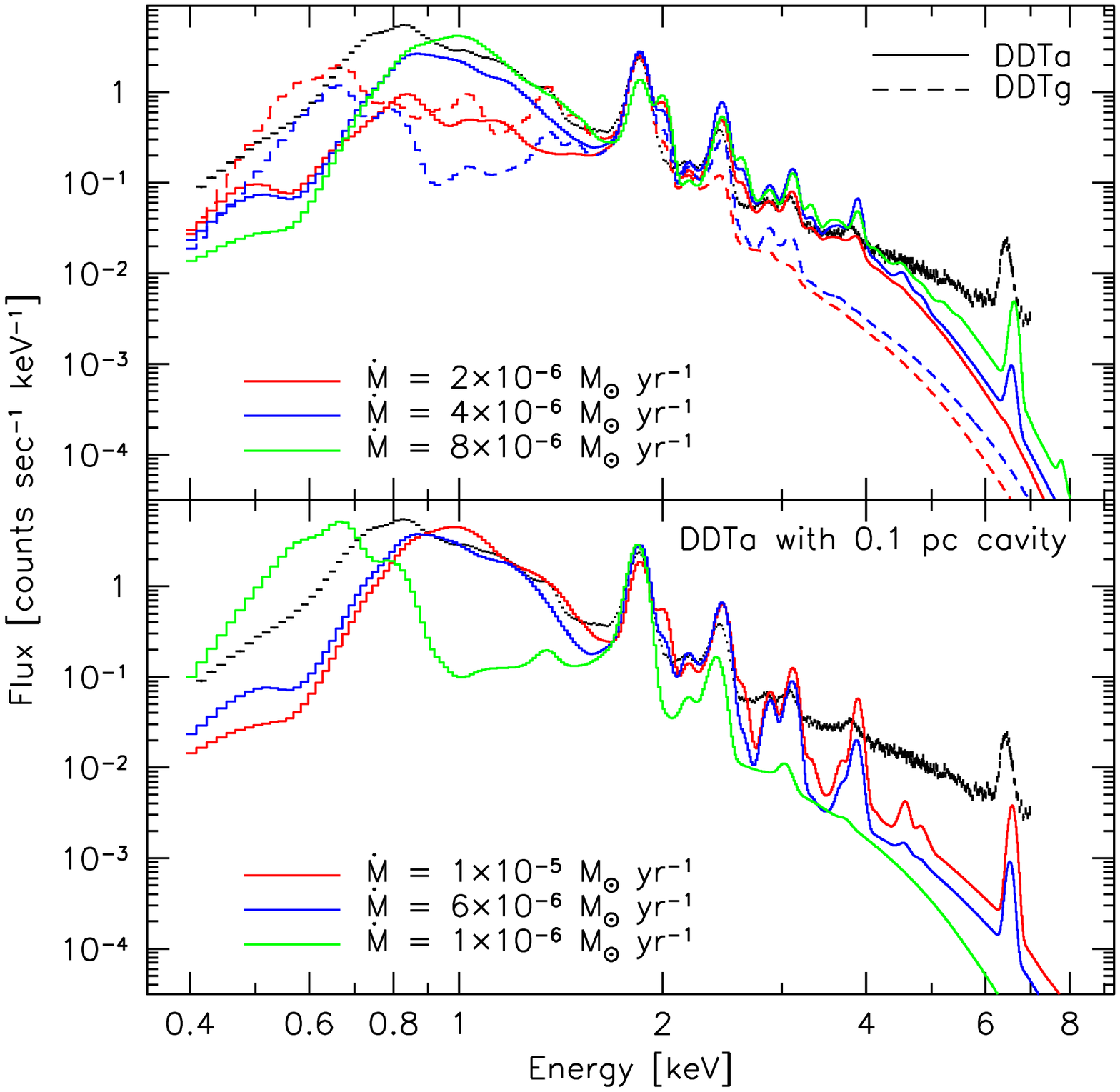}{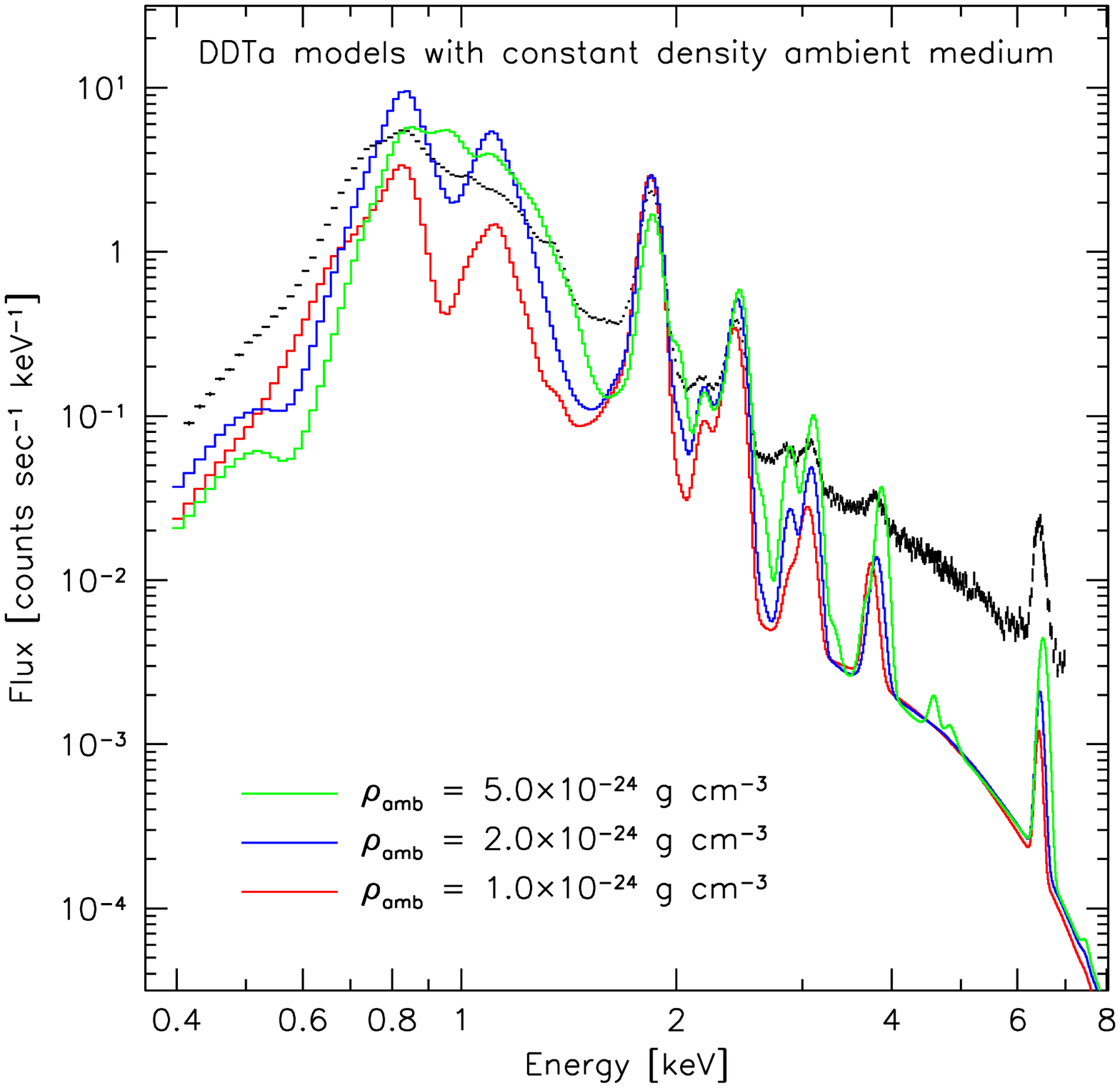}
\caption{{\it Left:} Comparisons between the modeled X-ray spectrum and the
spectrum shown in Figure~\ref{fig:kepler}. In the upper panel, we compare
the DDTa (solid lines) and DDTg (dashed lines) modeled spectra for a range
of mass-loss rates, assuming a wind speed of 20 km s$^{-1}$. In the bottom
panel, we plot the DDTa models for a wind speed of 20 km s$^{-1}$ and
a range of mass-loss rates, assuming a small, low density cavity has formed
around the progenitor. {\it Right}: DDTa models in a constant density 
ambient medium.}
\label{fig:emission}
\end{figure}

In Figure~\ref{fig:emission} (left, upper panel) we plot the 
X-ray spectrum from 
Figure~\ref{fig:spec_fit} (left) compared against a subset of our spectral models.
We plot the spectra from both the DDTa and DDTg models for a range
of mass-loss rates, assuming wind velocities of 20 km s$^{-1}$. Given
constraints (1) and (3) above, we can rule out the DDTg models since
they do not produce enough iron to match the emission seen both at 6.4 keV
and Fe L emission at $\sim$ 1 keV. 
By ruling out the DDTg models, we
are also essentially ruling out the distance estimate of 
3 $<$ D$_{\mathrm{SNR}}$ $<$ 6.4 kpc, if the blastwave is expanding 
into a $\rho_{amb}$ $\propto$ $r^{-2}$ wind. In contrast, some DDTa models do  
reproduce the Fe K$\alpha$ and Fe L emission.
Those DDTa models with with $\dot{M}$ $<$ 4$\times$10$^{-6}$ M$_{\sun}$
yr$^{-1}$ are ruled out since
they produce very little emission at 6.4 keV, in conflict with the data.

\begin{figure}
\plottwo{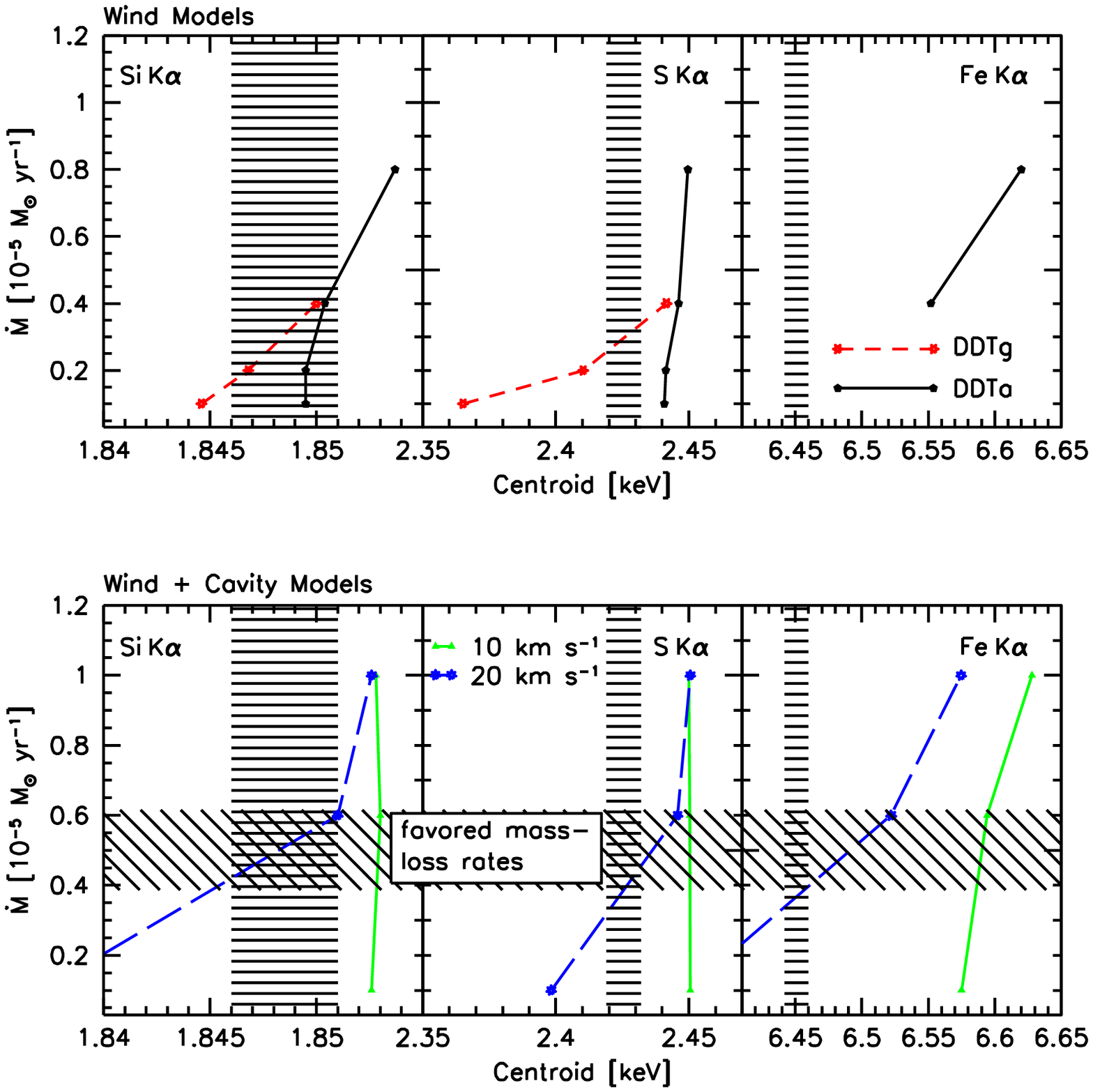}{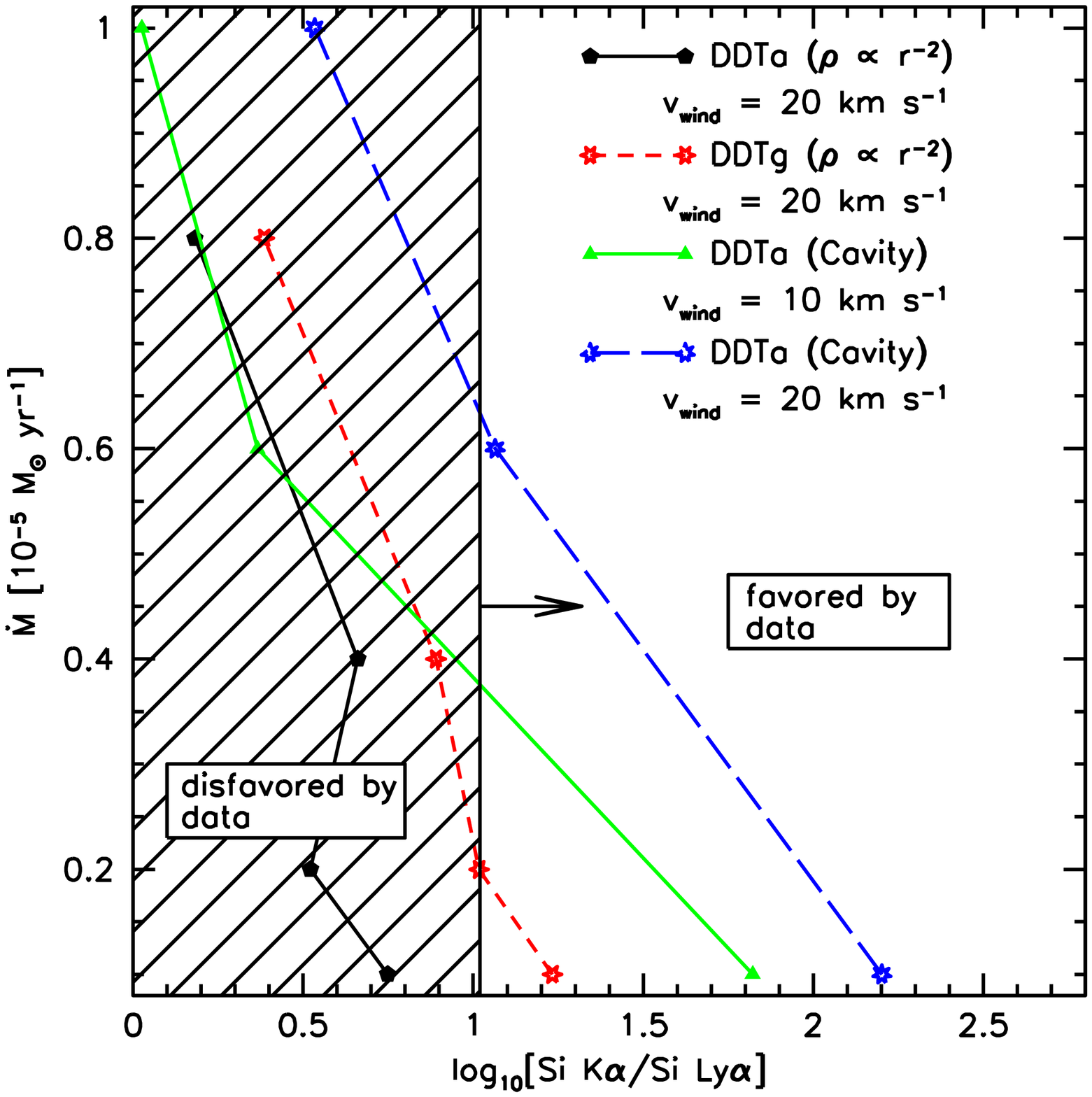}
\caption{{\it Left:} In the upper panel, we plot the measured
versus computed line centroids for Si K$\alpha$, S K$\alpha$, and Fe K$\alpha$, 
for the DDTa (black solid) and DDTg (red dashed) models. The hatched region in each panel
corresponds to the measured centroid including the 90\% confidence interval
The DDTg models and a subset of the DDTa models ($\dot{M}$ $<$ 4$\times$10$^{-6}$ 
M$_{\sun}$ yr$^{-1}$) do not produce any 
Fe K emission. In the lower panel, we plot the line centroids for the DDTa
cavity models for v$_{\mathrm{wind}}$ = 10 (green solid) and 20 (blue dashed) km 
s$^{-1}$, for a range of mass-loss rates. The line centroids and errors are indicated by 
the vertical hatched regions in each panel. The allowed mass-loss rates as
dictated by the comparison between the measured and modeled line centroids are 
marked by the horizontal cross hatched region.
{\it Right:} Si K$\alpha$/\ion{Si}{14} Ly$\alpha$ flux ratio from each model as compared
against the measured value from Table~\ref{tab:fits}. The flux ratios are plotted
as a function mass-loss rate for both the pure wind models as well as 
the models that include a cavity. The line styles are the same as in the left hand
plot. We mark the region of the plot that is not
consistent with the measured line ratio which rules out almost all pure wind models.}
\label{fig:cent_rat}
\end{figure}

In Figure~\ref{fig:cent_rat} (left, upper panels), we plot the
line centroids from the DDTa and DDTg models for various mass-loss rates
for wind speeds of 20 km s$^{-1}$. The hatched region in each panel corresponds to the 
90\% confidence on the line centroid from the spectral fitting in Table~\ref{tab:fits}. 
While we have previously ruled out the DDTg models in this wind scenario based
on the absence of Fe emission, we are also forced to rule out those DDTa models that
do produce Fe K emission
as the predicted line centroids for nearly all models considered are at 
higher energies than what is measured.

Finally, and for comparison with other possible CSM models, we note that those
DDTa models shown in Figure~\ref{fig:emission} that produce Fe K emission also result in 
significant \ion{Si}{14} Ly$\alpha$ emission at $\sim$ 2.0 keV. 
This emission is not observed in the spectrum (Fig.~\ref{fig:spec_fit}), and our
spectral fits only resulted in an upper limit on the \ion{Si}{14} Ly$\alpha$ flux 
(Table~\ref{tab:fits}). 
This is shown quantitatively in 
Figure~\ref{fig:cent_rat} (right) where we plot the ratio of the modeled flux from
Si K$\alpha$ to that from \ion{Si}{14} Ly$\alpha$, for comparison against the measured ratio.
In summary, based on the results of our comparisons between the observed and modeled 
bulk properties of the X-ray spectrum from the shocked ejecta,
we must conclude that the X-ray 
emission from Kepler's SNR is not consistent with what is expected from 
supernova ejecta interacting with a $\rho_{\mathrm{amb}}$ $\propto$ $r^{-2}$ 
wind as proposed by \citet{chiotellis12}. 

\subsection{Spectral Fitting: Wind + Cavity Model}

As discussed above, pure wind models do not provide a good fit to the
X-ray spectrum. The reason is that the low wind velocities and moderate
mass-loss rates that are required to match the SNR dynamics result in 
a significant amount of circumstellar material close to the explosion. 
Physically, 
this results in a stronger reverse shock being driven into the expanding
ejecta, resulting in higher average charge states-- this affects both the
K$\alpha$ line centroid as well as the amount of
Ly$\alpha$ emission. These are two measurable quantities which do not
agree with the simple models described above.

The possibility exists that the circumstellar environment is more
complicated than that of a simple $r^{-2}$ wind. Recurrent novae 
\citep[e.g.,][]{sokoloski06,woodvasey06} or optically
thick accretion winds \citep[e.g.,][]{hachisu96} from the white dwarf could 
carve out a cavity around the progenitor, even while the companion is losing
mass to a slow and dense wind.
We consider the formation of a small ($\sim$ 10$^{17}$ cm) cavity 
inside the
dense wind. The cavity could form in a number of ways \citep[see][for a recent 
review]{chomiuk12}, and time-variable \ion{Na}{1} D absorption spectra 
indicate that cavities can have sizes $\ga$ 10$^{17}$ cm 
\citep{patat07,blondin09,simon09,sternberg11}. A small cavity such as this 
will not significantly
alter the dynamics of the forward shock, but it will slow the ionization
of the ejecta, resulting in lower charge states and dynamically younger
shocked ejecta. 

In the bottom panel of Figure~\ref{fig:emission} (left), we plot
the synthesized X-ray emission from models where the wind has a small
cavity carved out of it. Additionally, we show in the bottom panel of 
Figure~\ref{fig:cent_rat}
(left) the computed line centroids for these models. 
In the right hand panel of
Figure~\ref{fig:cent_rat}, we show the modeled flux ratio of 
Si K$\alpha$ to \ion{Si}{14} Ly$\alpha$
compared against the upper limit on the flux ratio from the X-ray spectrum, as
a function of mass-loss rate. As shown
in these plots, the line centroids (Si K$\alpha$ and S K$\alpha$)
now agree with the measured values (for velocities $\lesssim$ 20 km s$^{-1}$).
Additionally, these modified models show little \ion{Si}{14} Ly$\alpha$
emission at $\sim$ 2 keV, as evidenced by the low flux ratios seen in
the right hand panel, particularly for mass-loss rates less than
6$\times$10$^{-6}$ M$_{\sun}$ yr$^{-1}$. We can thus conclude that these
models are consistent with the observed X-ray spectrum. 

When working within the dynamical constraints imposed by the DDTa models, we require
a distance greater than 7 kpc, and a wind luminosity greater than 
$\sim$ 6$\times$10$^{32}$ erg s$^{-1}$. Additionally, we require iron
emission, agreement between the modeled and measured K$\alpha$ line
centroids, and little or no \ion{Si}{14} Ly$\alpha$ emission. The high mechanical
wind luminosity rules out models with low values of $\dot{M}$, while the
lower limit on the distance rules out the high mass-loss rate models.
The presence of \ion{Si}{14} Ly$\alpha$ emission in the modeled spectra rule out those
models with $\dot{M}$ $\gtrsim$ 6$\times$10$^{-6}$ M$_{\sun}$ yr$^{-1}$.
We are left to conclude that a wind with a mass-loss rate of $\sim$ 4--6
$\times$10$^{-6}$ M$_{\sun}$ yr$^{-1}$ and wind velocities $\sim$ 20 km
s$^{-1}$ sculpted the CSM, and that a small cavity likely surrounded the 
progenitor, possibly caused by either recurrent novae or optically thick
accretion winds that preceded the supernova.

\subsection{Spectral Fitting: Constant Density Models}

Finally, we consider models with a uniform ambient medium, with density
$\rho_{\mathrm{amb}}$ = 1--5 $\times$ 10$^{-24}$ g cm$^{-3}$. \citet{vink08}
measured the expansion of Kepler's SNR in the south and compared it 
against models for the expansion of a SNR blastwave in a constant density
external medium. They found that models where $\rho_{\mathrm{ej}}$ $\propto$
$v^{-7}$ were able to reproduce the kinematics. However, they concluded
that the kinematics of the blastwave in the south point to an explosion
with E$_{\mathrm{SN}}$ $<$ 5 $\times$ 10$^{50}$ erg for the canonical 
distance of 4 kpc, assuming an ambient medium density of 
n$_{\mathrm{H}}$ = 1 cm$^{-3}$. 
Higher explosion energies work if the SNR is at a distance
considerably greater than 4 kpc, or if the ambient medium 
density $>$ 10 cm$^{-3}$.

\begin{figure}
\plotone{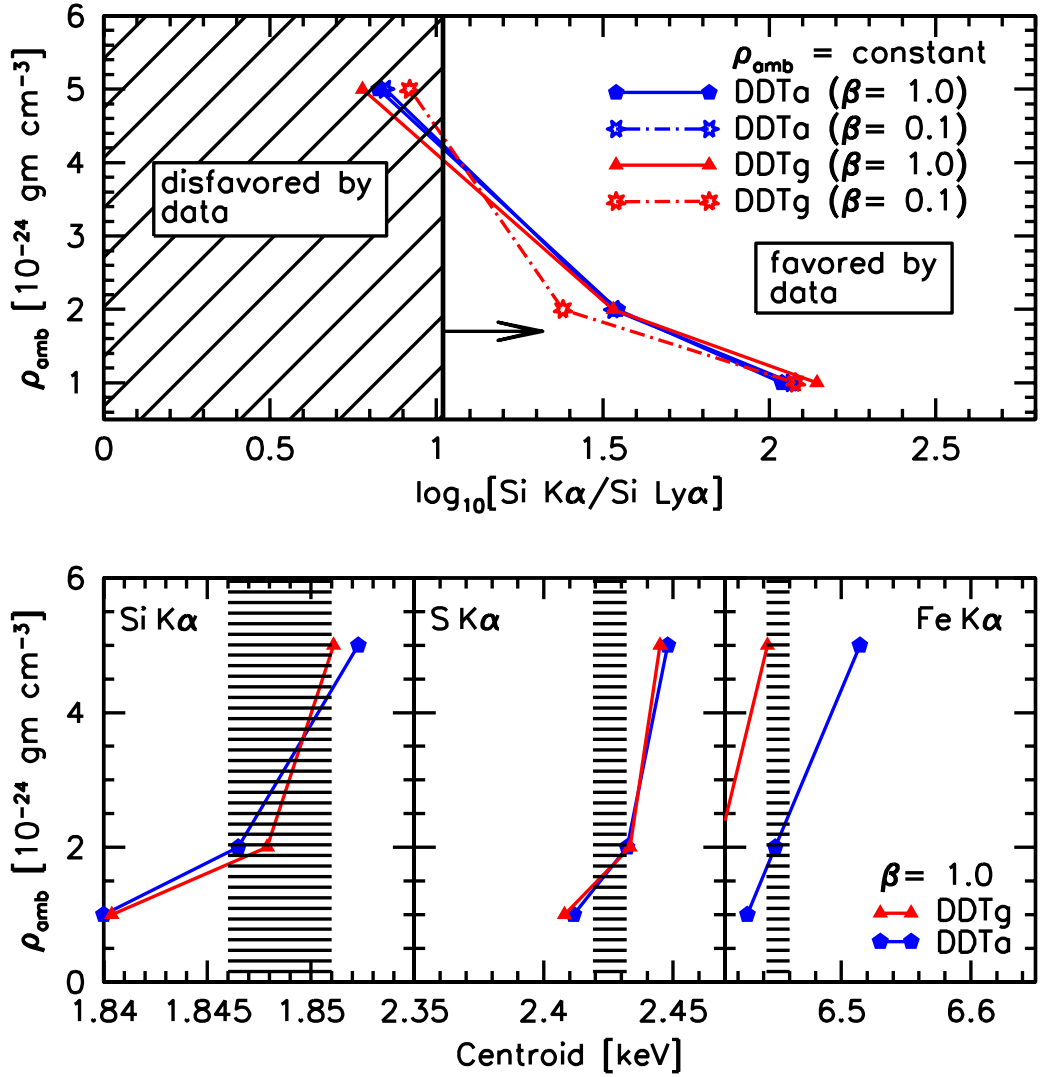}
\caption{{\it Top:} Si K$\alpha$/\ion{Si}{14} Ly$\alpha$ flux ratio for energetic
and subenergetic models in a constant density environment, for either
instantaneous ($\beta$ = 1) or mass proportional ($\beta$ = 0.1) 
electron heating. The modeled ratio is relatively
insensitive to the electron heating model. {\it Bottom:} Measured versus
computed line centroids for the DDTa (blue) and DDTg (red) models for
Si K$\alpha$, S K$\alpha$, and Fe K$\alpha$.}
\label{fig:const_models}
\end{figure}

In Figure~\ref{fig:emission} ({\it right}), we show the computed X-ray emission
for DDTa models in a constant density medium. As shown in 
Fig.~\ref{fig:emission}, these
models qualitatively reproduce the bulk features seen in the spectrum, 
and an ambient
medium density of 2--5 $\times$ 10$^{-24}$ g cm$^{-3}$ seems to be able to 
reproduce both
the line centroids as well as the line flux ratios. This is confirmed in 
Figure~\ref{fig:const_models}, where we see that models with densities
of $<$ 3$\times$10$^{-24}$ g cm$^{-3}$ produce the required \ion{Si}{14} Ly$\alpha$ 
emission and
also predict line centroids in agreement with the measured values. 
For these models, the forward shock radii at
400 yrs range from 2.9 -- 3.8 pc. For the measured angular radius of 
2$\arcmin$, this corresponds to a distance to Kepler of 5 -- 6.5 kpc,
consistent with the distance estimate of \citet{sankrit05}.

While the constant density models can explain the observed X-ray emission 
and also provide estimates for the distance to Kepler that are in agreement 
with proper motion measurements of Balmer filaments \citep{sankrit05}, 
it is difficult to reconcile a constant density CSM with the bowshock 
structure in the north. To form a bowshock, a converging flow is
required between an outflow from the progenitor and the bulk 
motion of the ambient ISM (in the rest frame of the progenitor). 
Two alternative scenarios are considered: 
First, for either a wind that is not spherically symmetric
but consists of equatorial and polar components \citep{owocki94}, or 
other non-spherical outflows, such as in the recurrent nova 
RS Oph \citep{sokoloski06,drake09,orlando09}, the CSM 
could be sufficiently modified such that it is better described as a constant
density environment, rather than a $r^{-2}$ wind. 

Alternatively, we can consider that Kepler's SNR exploded in an environment
with a density gradient. \citet{petruk99} modeled the evolution of a 
SNR in a large
density gradient. In Figure~5(a) of \citet{petruk99}, the surface brightness
profiles for the modeled remnant are qualitatively similar to the morphology of 
Kepler. That is, a bowshock like structure forms in the direction of the density
gradient. For RCW~86, \citet{petruk99} require a density contrast of $<$ 2 between the
north and south, in order to reproduce the observed X-ray emission at an age of
1800 yr. 
For Kepler, a density contrast considerably higher than 2 would be required
to produce the same morphology in just 400 yr. 160$\micron$ emission 
\citep{gomez12} suggests
a density gradient towards the north--northwest of Kepler, but they point out that
much of that emission {\it may} arise from foreground or background sources. \citep{blair07}
reached a similar conclusion from {\it Spitzer} 160$\micron$ observations, noting a
north--south gradient in the emission at that wavelength, indicative of a
density gradient in the circumstellar medium. If the bowshock structure is caused 
by a gradient in the density of the circumstellar environment, then that would suggest
that the SN exploded in an environment that was closer to uniform in density, rather
than one shaped by a slow and dense stellar wind interacting with the ISM. If
this were indeed the case for Kepler, then it would be similar to the candidate
Type Ia remnant G299.2-2.9. In the case of that remnant, a signficant density
gradient (3$\times$) is required to explain the observed 
asymmetry in the X-ray morphology \citep{park07}. However, that SNR is estimated
to be ten times older than Kepler, so the spatial scale of the density gradient
is correspondingly larger.

\section{Conclusions}

In order to explain the observed dynamics and X-ray emission of 
Kepler's supernova remnant, we have coupled models for ejecta from 
Type Ia SNe to a model for a circumstellar medium shaped by a slow stellar
wind. We have synthesized the X-ray emission from these models and
compared it to the observed X-ray spectrum. Our conclusions are:

\begin{enumerate}

\item If the distance to Kepler is taken to be less than 6.4 kpc, then the
dynamics (e.g. the forward shock radius) are better explained
by a subenergetic explosion model (DDTg) interacting with a slow and 
dense stellar wind. However, while the subenergetic models can explain the dynamics, 
the synthesized X-ray spectra do not agree with the observed X-ray emission. 
Specifically, the subenergetic model does not produce enough iron to explain 
the observed emission. The presence of strong Fe emission in the spectrum rules out all 
DDTg models.

\item If the distance is taken to be greater than 7 kpc, then the dynamics
can be explained by the energetic Ia models. While 
the X-ray spectrum of Kepler shows significant emission from Fe K and
Fe L, there is little emission from H-like Ly$\alpha$ of intermediate mass 
elements such as silicon. The modeled line centroids in the DDTa + wind models
do not agree with the measured line centroids, even in those models
with high mechanical luminosities (i.e. high wind speeds). Additionally, 
the pure wind models over predict the amount of \ion{Si}{14} Ly$\alpha$ emission relative
to that which is observed. This rules out the DDTa + wind models with CSM winds
that extend all the way to the explosion site.

\item Since the DDTg models are ruled out based on the presence of Fe emission, and
the DDTa + wind models are ruled out based on the measured K-shell line centroids and 
Si K$\alpha$ to \ion{Si}{14} Ly$\alpha$ flux ratio, we consider
the presence of a small, low density cavity around the progenitor prior to the SN.
The low density cavity alters the ionization state of the shocked ejecta 
but does not significantly
change the dynamics of the forward shock. We find that the presence of
the cavity reduces both the amount of \ion{Si}{14} Ly$\alpha$ emission as well as the
modeled line centroids to values consistent with the measured centroids.
We find that the energetic Ia model (DDTa) interacting with a wind with
velocity $\la$ 20 km s$^{-1}$ and mass-loss rate of $\la$ 6$\times$10$^{-6}$
M$_{\sun}$ yr$^{-1}$ is a good match both spectrally and dynamically,
assuming a distance $>$ 7 kpc.

\item DDTa models with a constant density ambient medium ($\rho_{\mathrm{amb}}$ $\approx$ 2
$\times$ 10$^{-24}$ g cm$^{-3}$) can explain the observed
X-ray emission and place the SNR at a distance that is consistent with previous
estimates \citep{sankrit05}. In order to explain the bowshock structure in the north, 
a north-south density gradient in the interstellar medium is required. Evidence for
such a gradient is seen in 160$\micron$ emission maps of Kepler and its vicinity
\citep{blair07}.\\

\end{enumerate}

Thus, we conclude that in order to explain both the dynamics and X-ray emission
of Kepler in the context of a wind environment, the explosion was energetic, producing
$\sim$ 1M$_{\sun}$ of $^{56}$Ni. The supernova occurred in a slow and dense CSM wind with a 
small central cavity.
The observed X-ray emission, which traces the explosion energetics, requires a
distance $\gtrsim$ 7 kpc, in the CSM wind model. Alternatively, we find that
the X-ray emission and dynamics of Kepler can also be explained by a constant CSM
density model. These models place the SNR at a distance of $\sim$ 5--6 kpc, consistent
with previous measurements. In the constant density scenario, a north--south 
density gradient in the ISM would be required in order to explain the SNR morphology. 

\acknowledgements

D.~J.~P. acknowledges support from the {\it Chandra} GO program through
grant GO0-11094X as well as support from NASA contract NAS8-03060. D.~J.~P. is
grateful for the hospitality of the Weizmann Institute for hosting him so that
the groundwork for this paper could be performed. We thank Alexa Hart at SAO for useful
discussions regarding the abundances and wind luminosities of asymptotic
giant branch stars. Additionally, we thank Paul Plucinsky for discussions regarding
the measurement of line centroids in ACIS spectra.

\begin{deluxetable}{lr}
\tablecolumns{2}
\tablewidth{0pc}
\tablecaption{Spectral Parameters in Kepler's SNR}
\tablehead{
\colhead{Parameter} & \colhead{Fitted Value\tablenotemark{a}}}
\startdata
\cutinhead{Power-Law Continuum}
$\alpha$ & 2.67$^{+0.06}_{-0.01}$\\
Norm (10$^{-3}$ photons cm$^{-2}$ s$^{-1}$ at 1 keV) & 1.90$^{+0.20}_{-0.02}$\\
\cutinhead{Line Fluxes (10$^{-5}$ photons cm$^{-2}$ s$^{-1}$)}
Si K$\alpha$\tablenotemark{b} & 49.9$\pm$0.20\\
\ion{Si}{14} Ly$\alpha$ & $<$ 4.70$\pm$0.47\\
Si K$\beta$  & 5.64$\pm$0.30\\
S K$\alpha$  & 13.5$\pm$0.30\\
Ar K$\alpha$ & 1.19$\pm$0.16\\
Ca K$\alpha$ & 0.38$\pm$0.07\\
Fe K$\alpha$ & 3.84$\pm$0.14\\
\cutinhead{Line Centroids (keV)}
Si K$\alpha$ & 1.848$_{-0.002}^{+0.003}$\\
Si K$\beta$  & 2.190$_{-0.003}^{+0.004}$\\
S K$\alpha$  & 2.425$_{-0.006}^{+0.007}$\\
Ar K$\alpha$ & 3.077$_{-0.006}^{+0.005}$\\
Ca K$\alpha$ & 3.799$_{-0.010}^{+0.010}$\\
Fe K$\alpha$ & 6.450$_{-0.008}^{+0.010}$\\
\cutinhead{Line Width (eV)}
Si K$\alpha$ & 25.9$_{-1.0}^{+0.5}$\\
Si K$\beta$  &  62.2$_{-3.2}^{+3.0}$\\
S K$\alpha$  &  35.9$_{-1.0}^{+1.5}$\\
Ar K$\alpha$ & 36.0$_{-5.5}^{+6.7}$\\
Ca K$\alpha$ & 57.9$_{-12.0}^{+13.0}$\\
Fe K$\alpha$ & 83.0$_{-5.1}^{+4.4}$\\
\enddata
\tablenotetext{a}{Errors correspond to 90\% confidence intervals}
\tablenotetext{b}{K$\alpha$ emission here refers to all allowed transitions to the ground state}
\label{tab:fits}
\end{deluxetable}

\end{document}